\documentclass[a4paper,11pt]{article}
\usepackage{pos}

\newcommand{\ON}{\textrm{O}(N)}
\newcommand{\calD}{\mathcal{D}}
\newcommand{\upd}{\textrm{d}}
\newcommand{\PT}{\mathcal{P}\mathcal{T}}
\newcommand{\ii}{\textrm{i}}
\newcommand{\LambdaC}{\Lambda_{\textsc{c}}}
\newcommand{\mubar}{\bar{\mu}}
\newcommand{\lambdaR}{\lambda_{\textsc{r}}}
\newcommand{\MR}{M_{\textsc{r}}}

\title{(In)stability of the Higgs vacuum from the $\textrm{O}(N)$ model at large $N$}

\author*[a]{Ryan Weller}
\author[a]{Chun-Wei Su}

\affiliation[a]{Department of Physics, University of Colorado at Boulder\\
  Boulder, CO, United States}

\emailAdd{ryan.weller@colorado.edu}

\abstract{The theory of an independent Higgs field is given by an $\textrm{O}(N)$ model with an $N$-component scalar $\vec{\phi}$ and a quartic $\lambda(\vec{\phi}\cdot\vec{\phi})^2$ potential when $N=4$. The phase structure of the theory can be studied analytically for all values of the coupling $\lambda$ using the large-$N$ limit, both at zero and finite temperature. However, authors in the 70s and 80s argued the theory at large $N$ was "sick" and "futile", and dismissed the theory. This was based on two points: (1) a failure to identify the stable phases and vacuum of the theory and (2) the issue of a negative bare coupling $\lambda<0$ in the UV. We provide evidence that the theory is not, in fact, "sick". Issue (2) is dealt with through the modern understanding of $\mathcal{P}\mathcal{T}$-symmetric non-Hermitian theories with "wrong-sign" couplings. Issue (1) is resolved by realizing that the true vacuum has no spontaneous symmetry breaking (SSB) and that the SSB phase only becomes preferred at high temperatures.}

\FullConference{42nd International Conference on High Energy Physics (ICHEP2024)\\
18-24 July 2024\\
Prague, Czech Republic\\}

\begin{document}
\maketitle

\section{Introduction}

Here we report on the thermodynamics and phase structure of the $\ON$ model in 3+1D at large $N$. The model's thermal partition function is given by a path integral
\begin{equation}
    Z=\int\calD^N\vec{\phi} \, e^{-S[\vec{\phi}]}
    \label{eq:part-func}
\end{equation}
with Euclidean action
\begin{equation}
    S[\vec{\phi}]=\int_{\beta,V} \upd^4 x \,\bigg(\frac{1}{2}(\partial_\mu\vec{\phi})^2+\frac{M^2}{2}\vec{\phi}^{\,2}+\frac{\lambda}{N}(\vec{\phi}^{\,2})^2 \bigg),
    \label{eq:action}
\end{equation}
with $\beta=1/T$ the inverse temperature, $M$ the bare mass and $\lambda$ the bare coupling. The $N$-component field $\vec{\phi}$ is a Lorentz scalar and an $\ON$ vector, hence this is sometimes called the ``vector model''. For $N=4$, this model is the theory of an independent Higgs field. Since the Higgs mechanism of the Standard Model relies on spontaneous symmetry breaking (SSB), it is of interest to study spontaneous symmetry breaking in the $\ON$ model.

The classic approach to studying this model is to do a perturbative expansion in small coupling $\lambda\ll1$, and argue that the $\ON$ symmetry is spontaneously broken based on the field living at the classical minima of the potential when $M^2<0$. This argument is well-motivated perturbatively, but may not be well-motivated non-perturbatively. Our approach is to do a calculation entirely non-perturbatively in the coupling, by employing the large-$N$ expansion, $N\gg1$. Large-$N$ techniques \cite{Romatschke-lecture} provide a useful alternative to the perturbative expansion. Results at large $N$ are valid for all values of the coupling $\lambda$ and give access to physics that would be otherwise ``unnoticed''. Moreover, renormalization at leading- and next-to-leading-order in $1/N$ is in some ways simpler than in perturbation theory \cite{Romatschke-renorm}. These techniques have been used to study equations of state \cite{Romatschke-asymptotic, Grable-Weiner} and bound states \cite{Romatschke-trivial}, and to calculate transport coefficients \cite{Aarts,Romatschke-analytic,Romatschke-shear,Lawrence-Romatschke,Weiner-Romatschke} like the shear-viscosity-to-entropy-density ratio $\eta/s$ and matter-to-curvature coupling $\kappa$ in scalar and fermionic field theories.

It was noticed in the 1970s that the $\ON$ model after large-$N$ renormalization has a negative bare coupling $\lambda<0$ \cite{Kobayashi}. Some authors dismissed the model as ``sick'' \cite{Coleman} because they found the SSB phase at low temperature was unstable. Thereafter, others noted that the stable phase at low temperature was an $\ON$-symmetric phase---that is, \textit{without} SSB---and asserted the model to be ``consistent'' \cite{Abbott}. But authors in the 1980s declared the theory to be ``futile'' and ``inconsistent'' \cite{Bardeen-Moshe-1} based on examining its high-$T$ behavior, and attributed this inconsistency to the negative bare coupling \cite{Bardeen-Moshe-2}. The presenter of this talk has likewise pointed out an apparent thermodynamic and dynamical instability at high temperature \cite{Weller}. 

We offer a possible solution by reexamining the phase structure and stability of the theory, including at high $T$, and arguing for an interpretation of the negative bare coupling $\lambda<0$. We reiterate that the SSB or ``standard Higgs'' vacuum is unstable and thermodynamically dispreferred at low $T$. We discuss the implication of this for Higgs physics, including for generating masses for gauge bosons in the Standard Model. 

\section{Results}

Results here are based on standard thermal field theory and large-$N$ techniques along the lines of \cite{Romatschke-lecture} coupled with the background field method for the field $\vec{\phi}$, which is split into a zero mode $\vec{\phi}_0$ plus non-zero modes. The Hubbard--Stratonovich auxiliary field $m^2=\ii\zeta$, which is the mass gap, is split into a zero mode $\ii\zeta_0$ plus non-zero modes, as is standard. The calculations will be presented in detail in forthcoming work \cite{Su}.

\subsection{Thermodynamics and phase structure}

In dimensional regularization, at leading order in large $N$, the renormalized coupling $\lambdaR(\mubar)$ and mass-squared $\MR^2(\mubar)$ as functions of the $\overline{\textrm{MS}}$ scale $\mubar$ satisfy
\begin{equation}
    \lambdaR(\mubar)=\frac{4\pi^2}{\ln(\LambdaC^2/\mubar^2)}, \,\,\,\,\,\,\,\,\,\,\,\,\,\,\,\frac{\MR^2(\mubar)}{\lambdaR(\mubar)} = \frac{\alpha\LambdaC^2}{4\pi^2},
    \label{eq:coupling}
\end{equation}
where $\LambdaC$ and $\alpha$ are two parameters associated with the theory. $\LambdaC$ is the dimensionful scale of the theory's Landau pole, as discussed in e.g. \cite{Romatschke-asymptotic, Weller}. The
resulting equation of state $P(T)$ (or pressure) of the theory at leading order in large $N$ is given by
\begin{equation}
    P(T)/N=p(T)=\frac{m^4}{64\pi^2}\bigg(\ln\bigg(\frac{\LambdaC^2}{m^2}\bigg)+\frac{3}{2}\bigg)-\frac{m^2}{8}\bigg(\frac{\alpha\LambdaC^2}{4\pi^2}+\frac{4}{N}\vec{\phi}_0^{\,2}\bigg)+\frac{m^2 T^2}{2\pi^2}\sum_{n=1}^{\infty} \frac{K_2(n\beta m)}{n^2}.
\end{equation}
Here, $m^2(T)=\ii\zeta_0(T)$ and the background field $\vec{\phi}_0(T)$ satisfy saddle conditions
\begin{equation}
    \frac{\partial p(T)}{\partial m^2}=0, \,\,\,\,\,\,\,\,\,\,\,\,\,\,\, \frac{\partial p(T)}{\partial \vec{\phi}_0}=0.
    \label{eq:saddles}
\end{equation}
At any given temperature $T$, only the saddles $m^2(T)$ and $\vec{\phi}_0(T)$ with the dominant pressure $p(T)$ contribute. There are phase transitions when two pressures $p_1(T)$ and $p_2(T)$ for two different solutions $\big(m^2(T),\, \vec{\phi}_{0}(T)\big)_1$ and $\big(m^2(T),\, \vec{\phi}_{0}(T)\big)_2$ to \eqref{eq:saddles} become equal or cross one another.

\begin{figure}
    \centering
 \includegraphics[width=.4\linewidth]{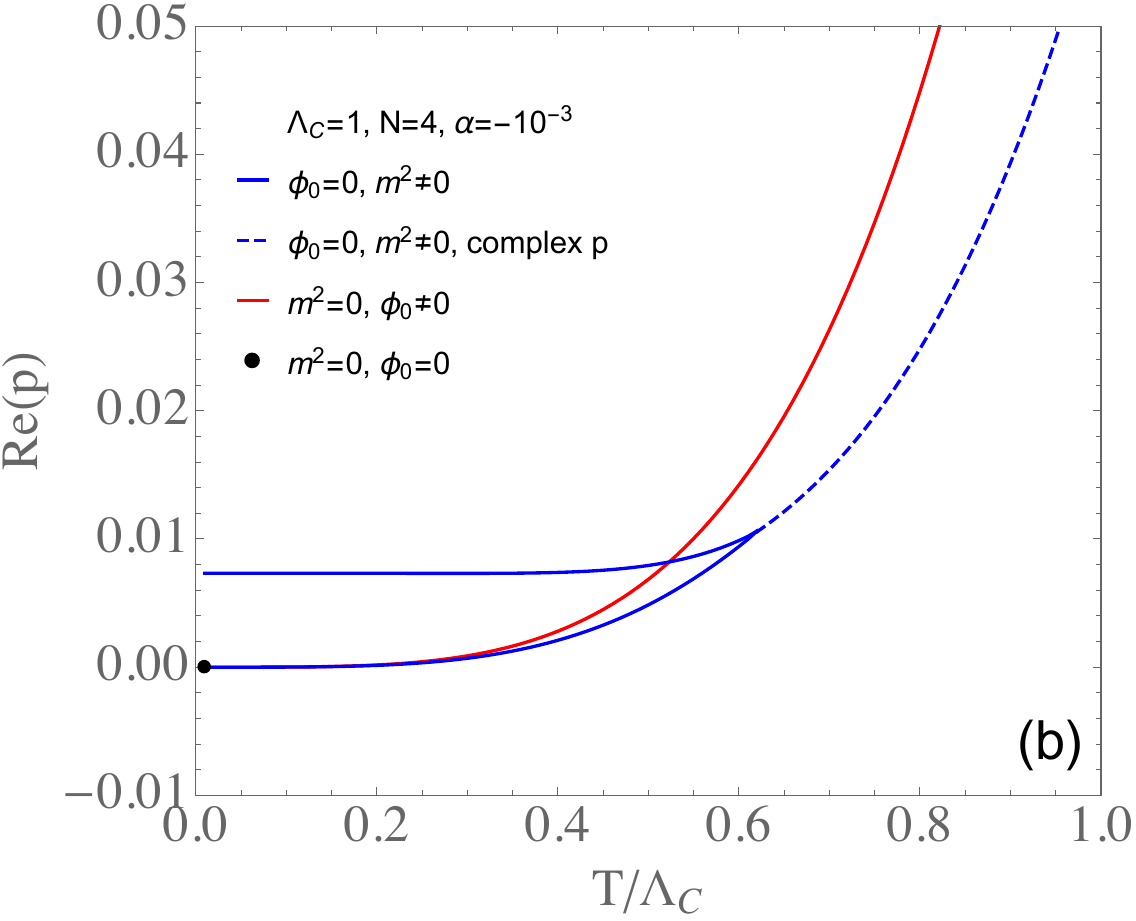}
\includegraphics[width=.4\linewidth]{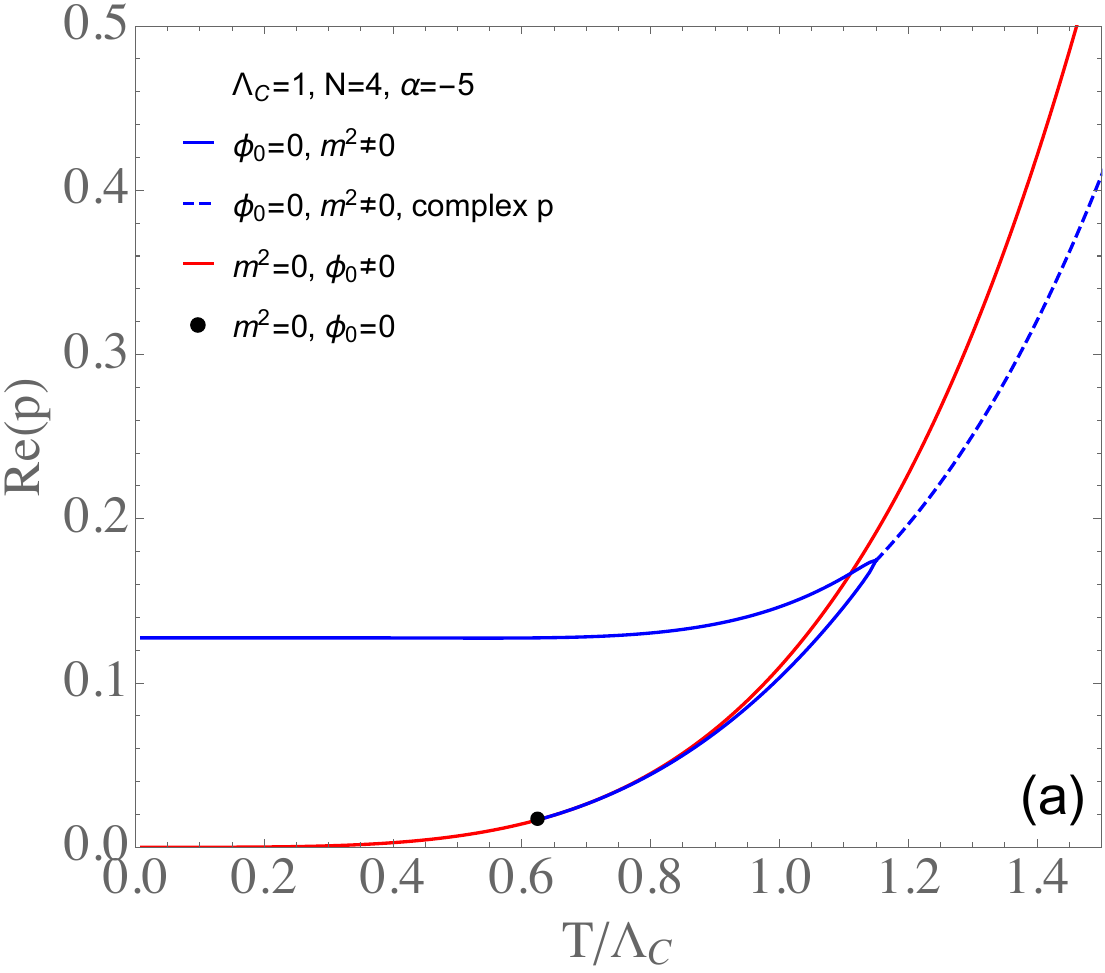}
    \caption{The pressure per component $p(T)/\LambdaC^4$ for two different values of $\alpha\propto M^2/\lambda$ as a function of temperature $T$ for the spontaneous-symmetry-broken phase (red) and the $\ON$-symmetric phases (blue) \cite{Su}. The dotted blue lines indicate a region where the pressures of the symmetric phases merge and become complex as discussed in \cite{Weller}. There is a first order phase transition where the red line crosses the upper blue line.}
    \label{fig:pressures}
\end{figure}

A plot of the pressure per component $p(T)$ for $\alpha=-5$ and $\alpha\approx 0$ is shown in figure \ref{fig:pressures}. For all values of $\alpha\lesssim0.824$ we find that the spontaneous-symmetry-broken phase $\vec{\phi}_0\neq 0$ (indicated in red) is \textit{not} thermodynamically preferred at low temperatures. This defies usual intuition, since one expects SSB perturbatively for $\alpha\propto M^2/\lambda<0$ at low temperatures. Hence the vacuum of the $\ON$ model for $\alpha\lesssim0.824$ is $\ON$-symmetric with $\vec{\phi}_0=0$, as known before \cite{Abbott}. It is curious to see, in a non-perturbative calculation, this great difference from what one would expect perturbatively. 

But at high temperatures, the SSB phase $\vec{\phi}_0\neq 0$ becomes preferred, with a pressure of a free massless boson $p(T)=\pi^2 T^4/90$. This too defies normal intuition, where SSB is expected only at low temperatures. One notes that the phase transition from SSB to a symmetric phase is first-order. This result might be of interest to cosmology, where a first-order phase transition in the early universe has observable implications e.g. in gravity waves \cite{Hindmarsh}. We will comment that the high-temperature SSB phase may only appear if we take the infinite volume limit $V\to\infty$ before the $m^2=0$ limit; this might have been missed by previous authors, but requires further study. Nicely, the SSB phase rescues the theory from the high-temperature thermodynamic instability reported by the presenter in \cite{Weller}. But it is unclear if this phase is dynamically stable, as we discuss in subsection \ref{subsec:stability}.

\subsection{Non-Hermiticity, $\lambda<0$, and $\PT$ symmetry}

 One sees that for scales $\mubar>\LambdaC$ above the Landau pole, the renormalized coupling $\lambdaR(\mubar)$ becomes negative, and therefore the bare coupling $\lambda$ is negative, $\lambda<0$, at high energies. Moreover, the pressure of the theory at high temperatures becomes that of $N$ non-interacting bosons, indicating asymptotic freedom of the theory. The asymptotic freedom is intimately connected to the negative UV coupling, as seen in the running coupling in \eqref{eq:coupling}, which goes to 0 from below as $\mubar\to\infty$. 
 
 The dominance of the ``free-theory-like'' SSB phase at high temperatures is curiously marked by a negative value of $\langle \vec{\phi}_0^{\,2} \rangle<0$. This suggests that the continuum limit of the renormalized $\ON$ model is a somehow a non-Hermitian theory. In fact, non-Hermiticity of the theory is required to make sense of the negative bare coupling $\lambda<0$, as pointed out in \cite{Romatschke-asymptotic,Romatschke-trivial} and as elaborated on by the presenter of this talk in \cite{Weller}.

 Such non-Hermitian theories, with negative or even complex couplings, can be well-behaved with real, lower-bounded energy spectra and a notion of unitarity, as long as they possess an antilinear symmetry, usually referred to as $\PT$ symmetry. This was first noted in the late 90s by Bender and Boettcher \cite{Bender}. They studied a class of Hamiltonians of the form $H=p^2-(\ii x)^K$ with the operator $x$ taking on eigenvalues in a complex domain. Therefore $x$ and $H$ in such theories are non-Hermitian operators. $\PT$ symmetry requires that the domain in which $x$ takes value is mapped onto itself under $\mathcal{P}: x\to-x$ and $\mathcal{T}:\ii\to-\ii$, and that the Hamiltonian $H$ respect this symmetry as well. In such a theory when $K=4$ (a negative quartic potential) it does indeed turn out that $\langle x^2 \rangle$ is negative in every energy eigenstate, similar to the negative $\langle \vec{\phi}_0^{\,2}\rangle<0$ we have observed at high $T$.

 Generalizing such a Hamiltonian to a quantum field theory, where $x(t)$ is promoted to a field $\phi(x)$, is an area of active research, e.g. \cite{Ai}. The presenter of this talk in \cite{Weller} has proposed that the continuum limit of the $\ON$ model is a $\PT$-symmetric theory with an ``upside down'' quartic potential $-g(\vec{\phi}^{\,2})^2/N$ for $g>0$, and where $\vec{\phi}$ is a non-Hermitian operator taking values in the domain
 \begin{equation}
     \vec{\phi}\cdot\hat{e}=\chi\big(\theta(\chi)e^{-\ii\pi/4}+\theta(\textrm{--}\chi)e^{\ii\pi/4}\big), \,\,\,\,\,\,\, \vec{\phi}-\hat{e}\vec{\phi}\cdot\hat{e}=\vec{\eta}\big(\theta(\chi)e^{-\ii\pi/4}+\theta(\textrm{--}\chi)e^{\ii\pi/4}\big),
 \end{equation}
 where $\hat{e}$ is a unit vector in $\mathbb{R}^N$ and where $\chi(x)\in\mathbb{R}$ and $\vec{\eta}(x)\in\mathbb{R}^{N-1}$ are real-valued fields parametrizing the complex domain. It appears to be showable  \cite{Weller-2} that on such a domain, after renormalization in 3+1D, observables are equivalent at leading order in large $N$ to those starting from the positive-coupling $\ON$ model defined by \eqref{eq:part-func}, \eqref{eq:action}. Thus a negative-coupling $\PT$-symmetric theory and the positive-coupling $\ON$ model seem to be ``equivalent'' at large $N$. This is rather strange and remarkable. In addition, this implies that the $\ON$ model at large $N$ has an interacting continuum limit as $\mubar\to\infty$ and is not quantum trivial \cite{Romatschke-trivial,Weller}. This is surprising, as scalar field theories  had been often expected to be trivial, i.e. to require a finite UV cutoff $\mubar<\LambdaC$ and have no interacting continuum limit. This kind of non-perturbative equivalence under renormalization, between an ``upside-down'' theory and a ``rightside-up'' theory, is also discussed in \cite{Grable-Weiner}. There is much more work to be done here.

\subsection{Dynamic stability of SSB phase}

\label{subsec:stability}

The $\ON$-symmetric phase at low temperatures can be shown to be dynamically stable \cite{Abbott} with a bound state excitation of the field $m^2=\ii\zeta$ that is not tachyonic \cite{Romatschke-asymptotic}.

We test the dynamic stability of the $\ON$-broken (SSB) phase by calculating the pole mass of the $\ii\zeta$ auxiliary field's propagator to leading order at large $N$. This requires calculating self-energies via so-called R2 resummation \cite{R2-1,R2-2,Romatschke-shear}. At $T=0$ there is a tachyonic excitation of the field $\ii\zeta$ for all $\alpha$. This indicates the zero-temperature dynamic instability of what is usually considered the ``Higgs vacuum'', and is consistent with what was noticed in the 70s \cite{Coleman}. At high temperatures, where the SSB phase dominates, however, there is no simple pole of the $\ii\zeta$ propagator, but rather a branch cut singularity. This is not classically a ``tachyon''; as of yet we have no word for it, and we are determining whether it is benign or renders the theory dynamically unstable at high $T$. More will be reported in \cite{Su}.

\subsection{Mass generation without SSB}

Lastly, given that the ``true'' ``Higgs vacuum'' at large $N$ is a vacuum without spontaneous symmetry breaking, it is reasonable to ask what should happen to the Higgs mechanism of the Standard Model. Without SSB, is it possible to generate masses for gauge bosons?

The answer to this question is ``yes''. It is possible using large-$N$ techniques to generate masses for a $\textrm{U}(1)$ gauge boson coupled to a complex scalar field $\vec{\phi}$ with $N+1$ independent components (see \cite{mass}). After gauge-fixing the Euclidean action of this theory is
\begin{equation}
        S[\vec{\phi},A]=\int_{\beta,V} \upd^4 x \,\bigg(\big(D_\mu\vec{\phi}\big)^*D_\mu\vec{\phi}+\frac{\lambda}{N}(\vec{\phi}^{\,2})^2 + (F_{\mu\nu}^2) \bigg),
\end{equation}
where $D_\mu=\partial_\mu-\ii e A_\mu /\sqrt{N}$ is the gauge-covariant derivative and $F_{\mu\nu}=\partial_\mu A_\nu-\partial_\nu A_\mu$ the field-strength tensor. Here after gauge-fixing, $\vec{\phi}$ is an $N$-component real scalar field.
The generation of mass for the gauge boson $A_\mu$ occurs in this model even though spontaneous symmetry breaking is not present. It happens even with a bare mass term $M^2=0$ of zero. Therefore it is possible to generate ``mass from nothing'', starting with no dimensionful parameters in the theory. The details of the calculation are in \cite{mass}.

In particular, in that work we estimate a Higgs-pair bound state with a mass at around $m_{\textrm{bound}}\approx400\textrm{ GeV}$ and a resonance at around $m_{\textrm{res}}\approx581\textrm{ GeV}$ and a width of $\Gamma_{\textrm{res}}\approx350\textrm{ GeV}$. These are in principle observable at the Large Hadron Collider (LHC), and an experimental search for these objects would be a test of the range of validity of the non-perturbative methods used here.

\section{Acknowledgments}
This work was in part supported by the Department of Energy, DOE award No. DESC0017905.

\end{document}